\documentclass[twocolumn]{elsart}

 \usepackage{epsfig}

\usepackage{amssymb}

\begin{document}

\begin{frontmatter}



\title{Determination of  $\gamma^{direct}/\pi^0$  using photon-charge particle correlation measurement in high energy heavy ion and pp collisions}


\author{Subhasis Chattopadhyay}

\address{Variable Energy Cyclotron Centre, 1/AF, Salt Lake , Calcutta-700064}

\begin{abstract}

A method has been proposed for the determination of $\gamma$ to $\pi^0$
 ratio in high energy pp and nucleus-nucleus collisions at large transverse
 momentum ($p_T$). In photon measuring
 device, it is proposed that shower shape analysis is made  to select only cases with wide showers.
 These candidates come mostly from $\pi^0$. Correlation is measured
between those photon candidates as trigger particles and associated charged
 particles above selected $p_T$.
 The ratio of yields of near and far angle correlation peaks
for $\pi^0$ events is compared with correlation peaks from other set of 
events where trigger particle contains photons for direct-photon jets alongwith $\pi^0$.
 The comparative study in reduction of yield of near angle peak is used to extract
 photon fraction.

\end{abstract}

\begin{keyword}
correlation, photon,jets
\PACS 
\end{keyword}
\end{frontmatter}

\section{Introduction}
\label{}

In high energy heavy ion collisions photons play a very important role
 in understanding the created system. Special attentions are made to
 detect photons at various $p_T$ \cite{photon}.
 Depending on the physics requirement, 
 energy and position resolution and energy reach of detected photons vary.
 At very high collision energy e.g at LHC a large fraction 
of photons are expected to
 be produced as $\gamma$-jet. 
Several observations are made at RHIC where leading particle spectra
are found to be quenched in high energy heavy ion collision.
 One of the possible explanation is loss
of energy of partons forming jets while passing through dense matter \cite{jet-quench}.
It is however impossible to reconstruct full jet in AA collisions 
because of large background due to soft particles.
 One of the ways calibration of jets can 
be done is by detecting photons from $\gamma$-jet events, where 
photons do not loose energy.
There are several predictions made on the cross-section of
 produced large $p_T$ photons where the fraction of photons
produced are expected to go up with increase in collision energy.
 On the other hand with increase in photon $p_T$,   the separation of photons
 from $\pi^0$ becomes difficult as the angular
separation between the decay photons become small and likely to deposit energy
in same detector cell. Efforts are made to reduce the cell-size of the
detector , so that some handles can be made on photon-$\pi^0$ discrimination
based on shower shape parameter. But even with reasonable granularity,
 photon samples contain
 good fraction of $\pi^0$ \cite{PHOS}. In these methods a threshold is applied on shower shape
 below which samples are rich in photons and at large shower-size
 the samples contain mostly $\pi^0$.
 These $\pi^0$-rich samples are usually ignored and efforts are made to
reduce the fraction of $\pi^0$ contamination in photon-rich sample.
 It is however possible to take these $\pi^0$-rich sample as reference and
 some $\pi^0$-specific property can be used to estimate $\gamma$ to $\pi^0$
 ratio in photon-rich sample. Overall $\gamma/\pi^0$ can be obtained 
taking $\gamma/\pi^0$ from various subset of events.\\
In this paper we propose photon-charged particle azimuthal
correlation for $\pi^0$-rich events as a reference
for such estimation. Decay photons from $\pi^0$ when correlated with
associated charged particles give two peak structures characteristics of
jets.  However, direct photons coming out as a part of photon jet will
have near angle peak absent and far-angle peak will look like
those of $\pi^0$-jet for similar jet energy.This property can be used to estimate photon-$\pi^0$ relative population.

\section{Proposal}
                                                                               
We propose that
we take photons with {\it very large shower size} as tigger particles
where samples contain mostly $\pi^0$ events 
and find the azimuthal correlation with respect to the associated charged 
particles above some $p_T$. 
We then measure the ratio ($R_{large}$) of the yields
 of two characteristic correlation peaks
by measuring the area under each correlation peak.
Similar correlation functions are measured for photons samples with
 {\it smaller shower
size} where data set contains $\gamma^{direct}$ with varying fraction.
 Here also we take the ratio of near and far angle yield ($R_{small}$).
Comparison of $R_{large}$ with $R_{small}$ will give the estimate of
$\gamma_{direct}$/$\pi^0$ for that set of events.
This exercise can be
carried out for different $p_T$ bins of trigger particles to obtain
$p_T$ spectra of $\gamma/\pi^0$.

                                                                               
\section{Simulation}
Simulation is performed using HIJING \cite{hijing} event generator in which hard-scattering
has been implemented with detailed jet fragmentation. No particular 
detector configuration is used in this simulation, so it is implied that 
 all particles produced in the events will be detected.
  We generated two types
 of events at LHC energy for pp collisions. One type of events select
only those cases having at-least one photon coming out from direct photon and
other type of events contain at least one high-$p_T$ $\pi^0$ leading particle.
Lowest threshold of $p_T$ for trigger photon
 used in this simulation is 8.5 GeV/c.
Correlation function is then constructed taking photons as trigger 
particle and charged particles with different $p_T$ as associated particle.
Fig.1 shows the correlation function for 'direct photon' events
 with $p_T^{associated} > 1$ GeV/c.
Fig 2 shows similar correlation function for $\pi^0$ events.
It is seen while in case of $\pi^0$ events two correlations peaks are clearly
 seen, in case of direct-photon events, near angle peak is absent. 
Large error bars in 'direct-photon' cases reflect lack of statistics.
In all correlation functions we measure (1/N$_{trig}$)dN/d$\phi$, 
where N$_{trig}$
is the number of trigger particle and dN corresponds to number of pairs in
 various $\Delta\phi$ bin. No correction is made on efficiency of associated
charged particles.
\begin{figure}[t]
\vbox{\hbox to\hsize{\psfig{figure=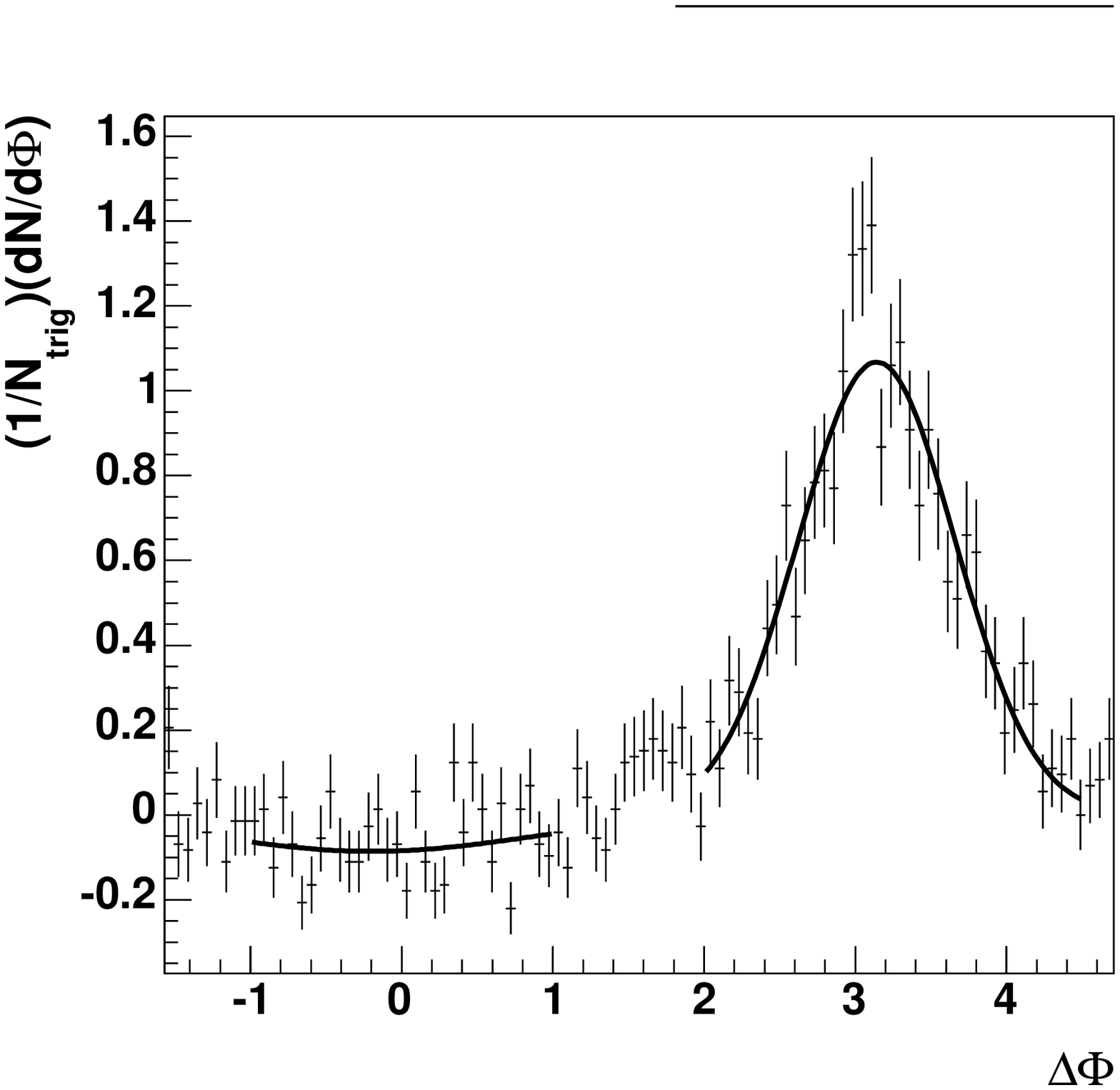,width=\hsize,clip=}\hfill}}
\caption{Azimuthal correlation for direct-photon events in pp collision at
LHC energy.
Photons with highest $p_T$ in the event is taken as trigger particle and 
charged particles with $p_T^{associated} > 1$ GeV/c is taken as
associated particles.
}
\end{figure}
                                                                               
\begin{figure}[t]
\vbox{\hbox to\hsize{\psfig{figure=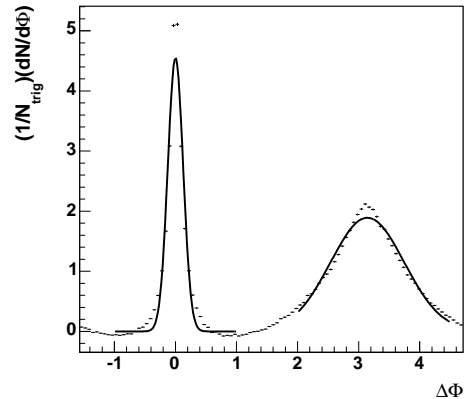,width=\hsize,clip=}\hfill}}
\caption{Azimuthal correlation for $\pi^0$ events in pp collision at
LHC energy.
Photons with highest $p_T$ in the event is taken as trigger particle and 
charged particles with $p_T^{associated} > 1$ GeV/c is taken as
associated particles.}

\end{figure}

Two peaks are then fitted with gaussian in the range of $\delta\phi$ = -1 to 1
 radians and $\delta\phi$ from 2.0 to 4.5 radians and yields are determined by
obtaining the area under each peak.
In order to simulate correlation function where $\pi^0$-jets are mixed with
$\gamma$-jets , we superposed direct-photon correlation
 function with
$\pi^0$-correlation function with different $\gamma/\pi^0$ ratios.
 Proper weightage is taken to obtain the resultant correlation function.
Fig. 3 shows the variation of (near angle yield)/(far angle yield) for
various $\gamma$-jet fraction in mixed sample.
 It is seen when entire data sample contains
$\gamma$-jet events, ratio becomes close to zero, and for various $\pi^0$
 fraction ratio increases reaching a value which depends on mean
 transverse energy of jets
and on associated particle threshold.
It is clearly seen that this yield ratio is sensitive to
 $\gamma$ to $\pi^0$
ratio.
\begin{figure}[t]
\vbox{\hbox to\hsize{\psfig{figure=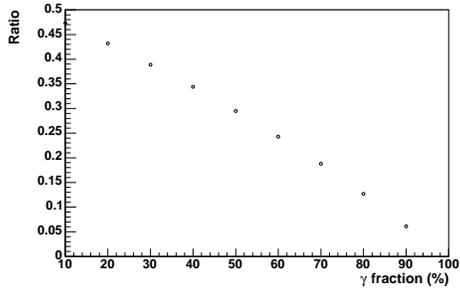,width=\hsize,clip=}\hfill}}
\caption{Ratio of near and far angle yield for various $\gamma$ fraction
 in the event samples where correlation function for direct-photon events
 are mixed with correlation function of $\pi^0$ leading particle events.}
\end{figure}
In order to simulate the sensitivity in case of PbPb collisions, we have
taken one sample PbPb event generated from Hijing event generator.
Two types of pp events are then superposed on this PbPb event.
Fig. 4 (top) shows the correlation function for the $\pi^0$-jet events
 superposed on PbPb event for $p_T^{trigger}>8.5$ GeV/c. As expected the
 background of the correlation function increases drastically in presence
 of PbPb event.
We then estimated the background by
 taking average of
 correlation function in the region of $\Delta\phi$ from 1 to 2 radians.
 Background was then subtracted to obtain final correlation function
(fig. 4(bottom) which
was then fitted with gaussians to obtain near and far angle
correlation function yield. 
 The peak heights in background-subtracted correlation function
 differ compared to pp case.
 This can be attributed
to the background in PbPb events where there are cases with
 large background not
 contributing to the correlation peak corresponding to the trigger particle.
But the method based on relative yield ratio should not be affected by this
 background effect. We therefore made all further investigations using background-subtracted correlation function.
Following the procedure as adopted for pp collision case, ratio is found for varying
$\gamma$-fraction.
It is observed that in case of correlation function in presence of heavy ion
 events also 
we can use this method to obtain $\gamma/\pi^0$. From the slope in two 
cases however it is seen that slope is steeper in pp case compared to 
PbPb case.
\begin{figure}[t]
\vbox{\hbox to\hsize{\psfig{figure=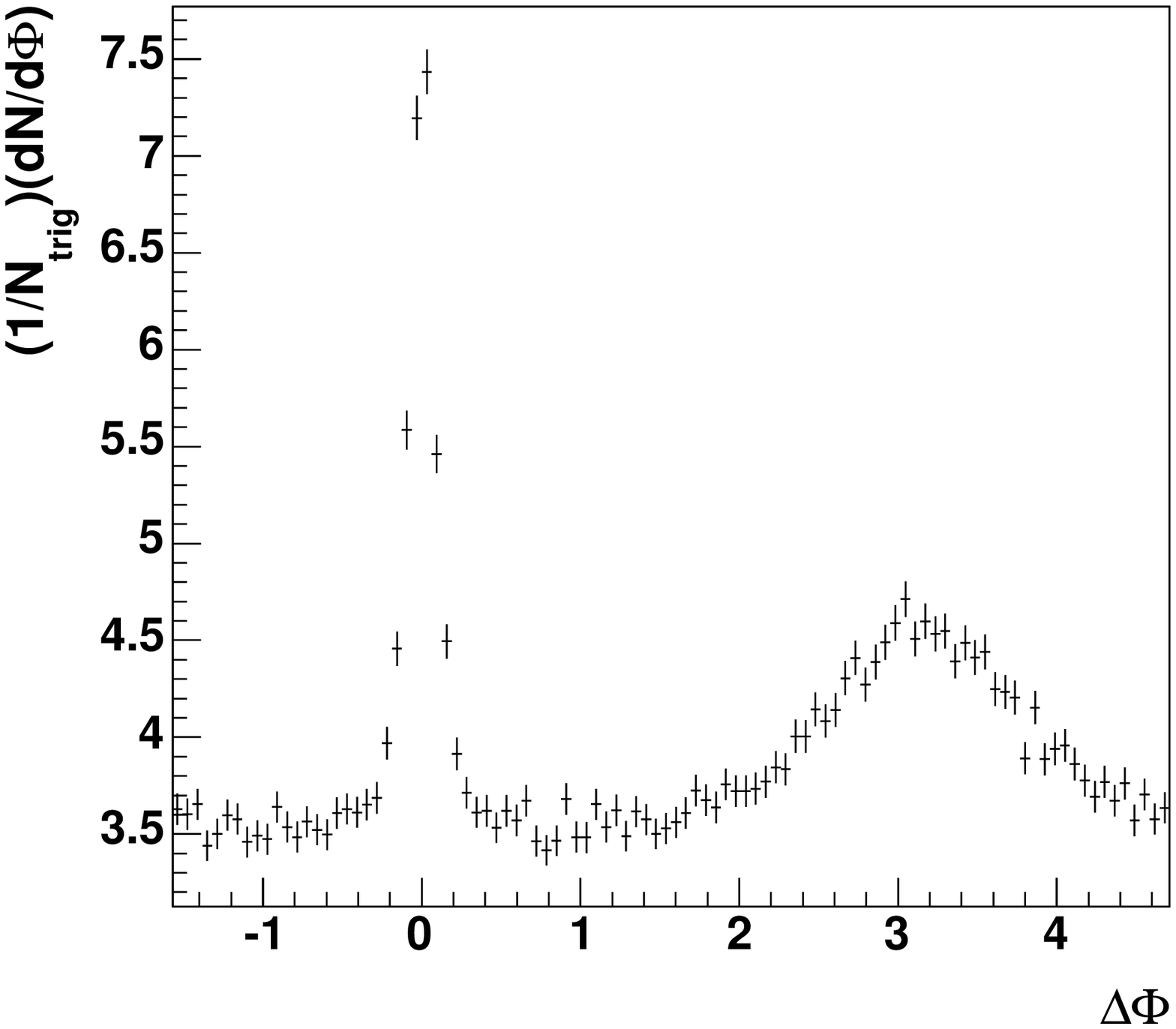,width=\hsize,clip=}\hfill}}
\vbox{\hbox to\hsize{\psfig{figure=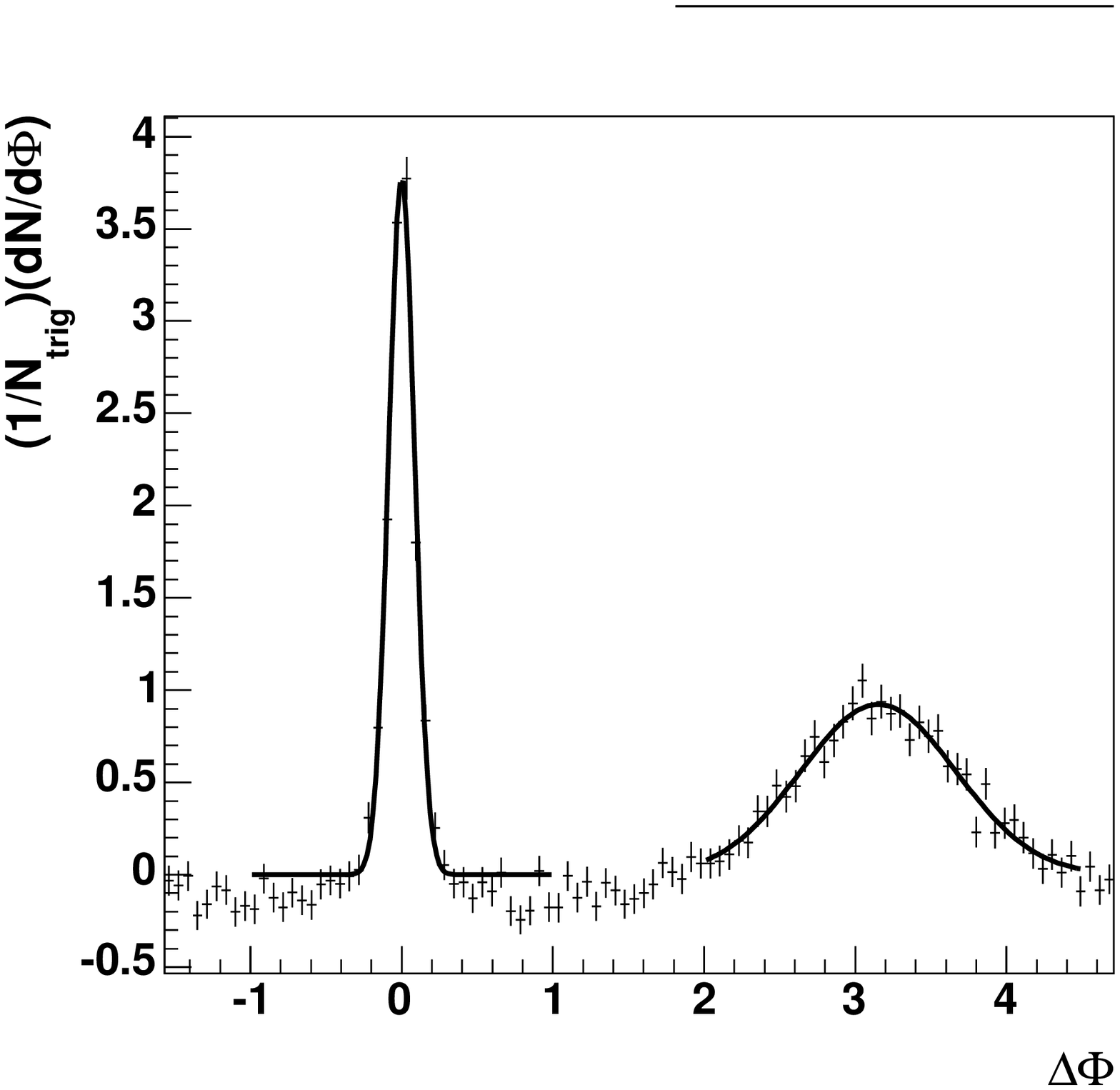,width=\hsize,clip=}\hfill}}
\caption{Azimuthal correlation function for $\pi^0$-jet
 pp events superposed on
 events from PbPb collisions (top) no background subtracted (bottom) background subtracted}
\end{figure}
                                                                                
\begin{figure}[t]
\vbox{\hbox to\hsize{\psfig{figure=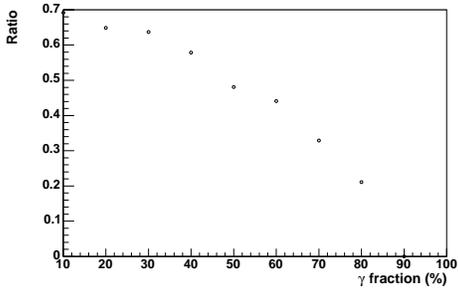,width=\hsize,clip=}\hfill}}
\caption{Ratio of near and far angle yield for various $\gamma$ fraction events
 superposed on PbPb events.}
\end{figure}
\section{Discussions}
It is clearly seen from the results of the simulation that more the sample
becomes rich in photon, better sensitive is the method in determining
 $\gamma/\pi^0$. Photon-rich samples can be obtained by two steps
 (a) using this currently proposed method after the use of all possible discriminatory
 properties available (b) we expect
increase in photon fraction due to the quenching of $\pi^0$.
There is however a possibility of mismatch for $\pi^0$-jet 
some fraction of energy going to the associated particle.
                                                                                
We have tried to estimate the effect of $p_T^{associated}$, as it might be
necessary to use higher threshold on $p_T^{associated}$ for PbPb central
events where background will be large. We have not made detailed estimation
 about the threshold of $E_T$ of trigger photons and corresponding $p_T^{associated}$ where correlation function can be meaningful, but we have taken
events with $E_T$ of trigger particle as 45-55 GeV/c and studied with three
 different $p_T^{associated}$ e.g. 1GeV/c, 2GeV/c and 4 GeV/c. For this
$E_T$ correlation peaks are clearly visible.
Fig.6 shows the ratio with different $\pi^0$ fraction, normalized with
the cases where photon fraction is absent.
It is seen that for lower $p_T^{associated}$, the ratio changes faster compared
 to higher $p_T^{associated}$. One of the possible reasons could be
with higher $p_T^{associated}$ full jet will not be accepted. So it might be
suggested to use as low $p_T^{threshold}$ as far as possible.
\begin{figure}[t]
\vbox{\hbox to\hsize{\psfig{figure=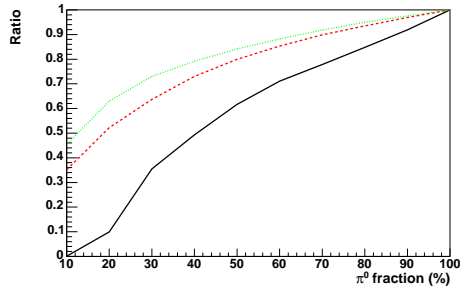,width=\hsize,clip=}\hfill}}
\caption{Ratio of near and far angle yield for various $\pi^0$ fraction for
 $p_T^{trig} > 45-55$ GeV/c and $p_T^{associated} > 1, 2, 4$ GeV/c.
  Ratio for each case is normalized wth respect to
 the ratio when entire sample contains $\pi^0$ events. Solid line correspond
 to $p_T^{associated} > 1$ GeV/c, dashed line for $p_T^{associated} > 2$ GeV/c
 and dotted line correspond to $p_T^{associated} > 4$ GeV/c. }
\end{figure}
                                                                                
This approach are expected to be useful in extracting the spectra of photons
after the estimation of $\pi^0$ production using $\pi^0$-rich sample.
This estimation of the sensitivity of the method however should be made using realistic photon production
rate at LHC energies.

In summary, we have proposed a method for extraction of $\gamma/\pi^0$ for
 heavy ion collisions at LHC energies using photon-charged correlation.
It is seen that the yield ratio of near and far angle correlation peaks
is sensitive to $\gamma/\pi^0$ in the sample of events. It is also seen that
this method is more effective for lower $p_T^{associated}$.



\end{document}